\documentclass[twocolumn,english,pra]{revtex4}
\usepackage[T1]{fontenc}
\usepackage[latin9]{inputenc}
\usepackage{subfigure}
\usepackage{amsbsy}
\usepackage{graphicx}
\usepackage{amssymb}
\usepackage{esint}

\makeatletter

\newcommand{\noun}[1]{\textsc{#1}}
\providecommand{\tabularnewline}{\\}

\usepackage{psfrag}
\makeatother

\usepackage{babel}

\begin{document}

\title{Motional Spin Relaxation in Large Electric Fields}

\author{\noun{Riccardo Schmid, B. Plaster, B. W. Filippone}}

\affiliation{California Institute of Technology, Pasadena, CA 91125}

\begin{abstract}
We discuss the precession of spin-polarized Ultra Cold Neutrons (UCN)
and $^{3}$He atoms in uniform and static magnetic and electric fields
and calculate the spin relaxation effects from motional $v\times E$
magnetic fields. Particle motion in an electric field creates a motional
$v\times E$ magnetic field, which when combined with collisions,
produces variations of the total magnetic field and results in spin
relaxation of neutron and $^{3}$He samples. The spin relaxation times
$T_{1}$ (longitudinal) and $T_{2}$ (transverse) of spin-polarized
UCN and $^{3}$He atoms are important considerations in a new search
for the neutron Electric Dipole Moment at the SNS \emph{nEDM} experiment.
We use a Monte Carlo approach to simulate the relaxation of spins
due to the motional $v\times E$ field for UCN and for $^{3}$He atoms
at temperatures below $600\,\mathrm{mK}$. We find the relaxation
times for the neutron due to the $v\times E$ effect to be long compared
to the neutron lifetime, while the $^{3}$He relaxation times may
be important for the \emph{nEDM} experiment.
\end{abstract}
\maketitle

\section{Introduction}

\noindent The phenomenon of longitudinal and transverse spin relaxation
due to magnetic field inhomogeneities has been discussed extensively\citep{PhysRev.138.A946,PhysRev.139.A1398,PhysRevA.37.2877,PhysRevA.38.5092,McGregor1990}.
There, the fluctuating magnetic fields experienced by the particle
spins during transport through magnetic field gradients coupled to
collisions with other particles or a holding cell's walls lead to
relaxation. For a combined magnetic and electric field configuration,
another possible source of spin relaxation has been considered by
\citet{PhysRevA.53.R3705} and results from fluctuations in the motional
magnetic field in the particle's rest frame. Here we present a more
detailed discussion of this effect.

\begin{figure}[b]
\subfigure[]{\includegraphics[scale=0.5]{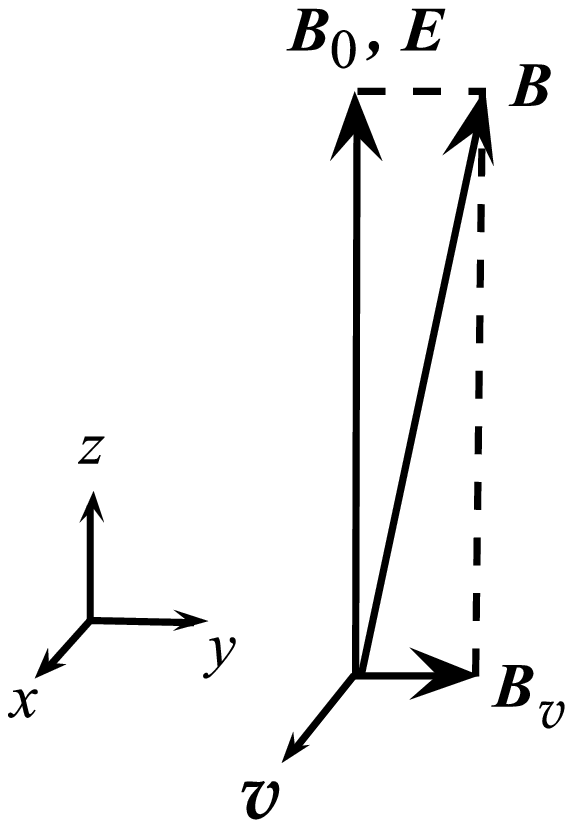}}\subfigure[]{\includegraphics[scale=0.5]{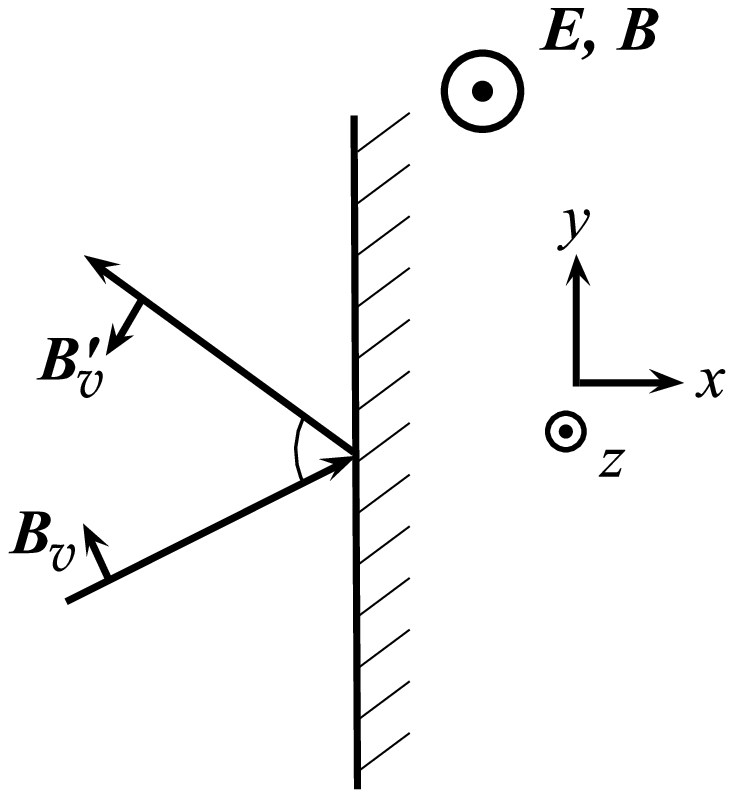}}

\caption{\label{fig:vxEfield}(a) The total magnetic field $B$ as the superposition
of the holding field $B_{0}$ and the $v\times E$ motional field
$B_{v}$. (b) Rotation of motional field $B_{v}$ due to wall collision.}

\end{figure}

\noindent Transformation into the rest frame of a particle moving
in an electric field gives rise to a motional magnetic field\begin{equation}
\boldsymbol{B_{v}}=\boldsymbol{E}\times\boldsymbol{v}/c^{2}.\end{equation}
The motional magnetic field is always perpendicular to both the direction
of motion and the electric field. For parallel electric and magnetic
field configurations (as are typically employed in particle electric
dipole moment searches), the superposition of $\boldsymbol{B_{v}}$
and a magnetic holding field $\boldsymbol{B_{0}}$ leads to a small
tilt in the total magnetic field $\boldsymbol{B}=\boldsymbol{B_{0}}+\boldsymbol{B_{v}}$
(see Fig. \ref{fig:vxEfield}), as the motional magnetic fields are
typically very small relative to $B_{0}$. For example, in an upcoming
search (\emph{nEDM})\citep{nedm} for the neutron electric dipole
moment at the Spallation Neutron Source (SNS) in the Oak Ridge National
Laboratory, the motional magnetic field experienced by the neutrons
from the $50\,\mathrm{kV/cm}$ electric field will be on the order
of $5.6\times10^{-4}\,\textrm{mGauss}$ per m/s of velocity, whereas
the magnetic holding field $B_{0}$ will be about $10\,\mathrm{mGauss}$. 

Nevertheless, collisions result in an abrupt change in the velocity
vector, causing the motional magnetic field $\boldsymbol{B_{v}}$
to change both in direction and magnitude. Although the changes in
orientation of the total magnetic field following a single collision
are rather small (as $B_{v}\ll B_{0}$), after each collision, a particle's
spin will precess about a slightly reoriented total magnetic field.
After a large number of such collisions, each of these small {}``kicks''
in the field will contribute to the overall global relaxation of a
spin ensemble.

In the remainder of this paper, we calculate the magnitude of spin
relaxation due to the $v\times E$ effect, especially as relevant
for the conditions of the upcoming \emph{nEDM} search, which will
utilize Ultra Cold Neutrons (UCN, neutrons with speeds $<7\,\mathrm{m/s}$)
and an in-situ polarized $^{3}$He {}``co-magnetometer'' immersed
in a superfluid $^{4}$He bath at temperatures in the range $100-600\,\mathrm{mK}$.
Spin-relaxation times significantly longer than the neutron lifetime
(limiting the measurement time) are desired.

The results in this paper are complementary to the studies on spin
relaxation due to magnetic field gradients\citep{PhysRev.138.A946,PhysRev.139.A1398,PhysRevA.37.2877,PhysRevA.38.5092,McGregor1990}
and calculations of the linear electric field geometric phase effects\citep{geometricPhasePendlebury,LamoreauxGolub2005,geometricPhase},
both of which can be important considerations for the measurement
of EDMs.

\section{Theory\label{sec:Theory}}

\subsection{$^{3}$He Atoms\label{sub:Theory-3He-Atoms}}

We have followed the formalism of \citet{McGregor1990} to describe
the relaxation process of the $v\times E$ motional field. This uses
the field auto-correlation function to calculate longitudinal ($T_{1}$)
and transverse ($T_{2}$) spin relaxation times. The spin relaxation
times for particle spins due to a changing $B$ field can be calculated
using the field auto-correlation function. With this method\citep{McGregor1990},
the longitudinal spin relaxation rate can be expressed as \begin{equation}
\frac{1}{T_{1}}=\frac{\gamma^{2}}{2}\left[S_{Bx}(\omega_{0})+S_{By}(\omega_{0})\right],\label{eq:McGdepFormula}\end{equation}
where $\gamma$ is the gyromagnetic ratio and $S_{Bx}(\omega_{0})$
and $S_{By}(\omega_{0})$ are the Fourier transforms of the field
auto-correlation function. For a perturbing field $\boldsymbol{\widetilde{B}}$,
the frequency spectrum is given by \begin{equation}
S_{B_{i}}(\omega)=\int_{-\infty}^{\infty}\left\langle \widetilde{B}_{i}(t)\,\widetilde{B}_{i}(t+\tau)\right\rangle e^{-\imath\omega\tau}d\tau\label{eq:Spectrum}\end{equation}
 for each of the fluctuating field components. The average inside
the integral in Eq. \ref{eq:Spectrum} is calculated for particles
in the measurement cell colliding with the walls of the cell and with
excitations in the superfluid $^{4}$He.

For the case of particles in a uniform magnetic holding field in the
$\hat{\boldsymbol{z}}$ direction and moving at velocity $\boldsymbol{v}$
in a uniform electric field $\boldsymbol{E}$ parallel to $\boldsymbol{B}_{0}$,
the non-zero perturbation magnetic field terms are\begin{equation}
\begin{array}{rcl}
\widetilde{B}_{x} & = & -E\, v_{y}/c^{2}\\
\widetilde{B}_{y} & = & E\, v_{x}/c^{2}\end{array}.\end{equation}
The field auto-correlation function, for this case, is proportional
to the velocity auto-correlation function, as \begin{equation}
\begin{array}{rcl}
\left\langle \widetilde{B}_{x}(t)\,\widetilde{B}_{x}(t+\tau)\right\rangle  & = & \left\langle v_{y}(t)\, v_{y}(t+\tau)\right\rangle \, E^{2}/c^{4}\\
\left\langle \widetilde{B}_{y}(t)\,\widetilde{B}_{y}(t+\tau)\right\rangle  & = & \left\langle v_{x}(t)\, v_{x}(t+\tau)\right\rangle \, E^{2}/c^{4}\end{array}.\end{equation}
For this work we adopt the velocity auto-correlation function\citep{McGregor1990}
\begin{equation}
\left\langle v_{x}(t)\, v_{x}(t+\tau)\right\rangle =\left\langle v_{x}^{2}\right\rangle e^{-\tau/\tau_{c}},\label{eq:v_autocorrelation}\end{equation}
where $\tau_{c}$ is the mean time between collisions. Eq. \ref{eq:v_autocorrelation}
is applicable in determining the frequency spectrum when the Larmor
frequency $\omega_{0}$ is larger than $1/\tau_{D}$, where $\tau_{D}$
is the diffusion time. This is the case for $^{3}$He atoms at temperatures
$T>400\,\mathrm{mK}$, where, as discussed in Sec. \ref{sec:He3 atoms},
collisions with excitations in the superfluid are frequent and atoms
take a relatively long time to travel to the walls of the holding
cell. The corresponding field auto-correlation frequency spectrum
is \begin{equation}
S_{Bx}(\omega)=\frac{2\left\langle v_{y}^{2}\right\rangle \tau_{c}}{1+\omega^{2}\tau_{c}^{2}}\,\frac{E^{2}}{c^{4}},\label{eq:spectrum_derived}\end{equation}
and a similar formula for $S_{By}(\omega)$. With Eq. \ref{eq:McGdepFormula},
we find the longitudinal spin relaxation time to be\begin{equation}
T_{1}=\frac{3}{2}\,\frac{1}{\gamma^{2}}\,\frac{c^{4}}{E^{2}\, v_{rms}^{2}\,\tau_{c}}\,\left(1+\omega_{0}^{2}\,\tau_{c}^{2}\right),\label{eq:McGregorT1}\end{equation}
 where we have used $\left\langle v_{x}^{2}\right\rangle +\left\langle v_{y}^{2}\right\rangle =2\, v_{rms}^{2}/3$.
As discussed in Sec. \ref{sec:He3 atoms}, the mean time between collisions
drops dramatically as the temperature of the $^{3}$He sample increases,
and, in the \emph{nEDM} setting, $\omega_{0}\tau_{c}\ll1$ for $T>300\,$mK,
so that Eq. \ref{eq:McGregorT1} can be approximated to be \begin{equation}
T_{1}\simeq\frac{3}{2}\,\frac{1}{\gamma^{2}}\,\frac{c^{4}}{E^{2}\, v_{rms}^{2}\,\tau_{c}},\end{equation}
leaving the relaxation times independent of the magnetic holding field
$B_{0}$ for this regime.

The transverse spin relaxation rate can be expressed as\citep{McGregor1990}\begin{equation}
\frac{1}{T_{2}}=\frac{1}{2T_{1}}+\frac{\gamma^{2}}{2}S_{Bz}(0),\end{equation}
but, for the case of $v\times E$ fields in which the perturbation
field has no z-component, the term $S_{Bz}(0)=0$, so we find that
$T_{2}=2T_{1}$.

\subsection{Ultra Cold Neutrons\label{sub:Theory-UCN}}

Although Eq. \ref{eq:McGdepFormula} can be applied to any perturbative
field, as long as the proper expression for the auto-correlation function
is used, we note that the earlier formalism detailed in \citet{PhysRev.138.A946}
is useful for the special case of UCN whose velocities do not change
in the \emph{nEDM} measurement cells. Following their procedure to
calculate the spin relaxation in inhomogeneous fields, which does
not rely on the field auto-correlation function, the expression for
the longitudinal relaxation time $T_{1}$ due to motional $v\times E$
field can be calculated to be\begin{equation}
T_{1}=\frac{3}{4}\cdot\frac{(1.19)^{2}\lambda\, B_{0}^{2}\, c^{4}}{v^{3}\, E^{2}}\cdot\frac{1+(\omega_{0}\tau_{c})^{2}}{(\omega_{0}\tau_{c})^{2}}.\label{eq:Theory}\end{equation}
This formula is equivalent to Eq. \ref{eq:McGregorT1}, apart from
small constant terms. It is also interesting to note that, for the
case of UCN in the \emph{nEDM} experiment, the mean free path $\lambda$
in Eq. \ref{eq:Theory} is a constant of the geometry of the measurement
cell, since UCN do not interact with other particles. In addition,
$\omega_{0}\tau_{c}>1$ for UCN speeds, so that the last factor in
Eq. \ref{eq:Theory} is roughly constant and of order unity. In this
case, Eq. \ref{eq:Theory} can be approximated to be \begin{equation}
T_{1}\simeq\frac{3}{4}\cdot\frac{(1.19)^{2}\lambda\, B_{0}^{2}\, c^{4}}{v^{3}\, E^{2}},\end{equation}
emphasizing the $T_{1}\propto1/v^{3}$ relation. The relaxation time
can be lengthened by increasing the magnitude of the magnetic holding
field $B_{0}$. However, as seen in Sec. \ref{sec:Ultra-Cold-Neutrons},
the motional spin relaxation times for UCN in the \emph{nEDM} experiment
are long compared to the neutron lifetime.

\section{\label{sec:Simulation}Simulation}

\begin{figure}
\includegraphics[width=3.4in]{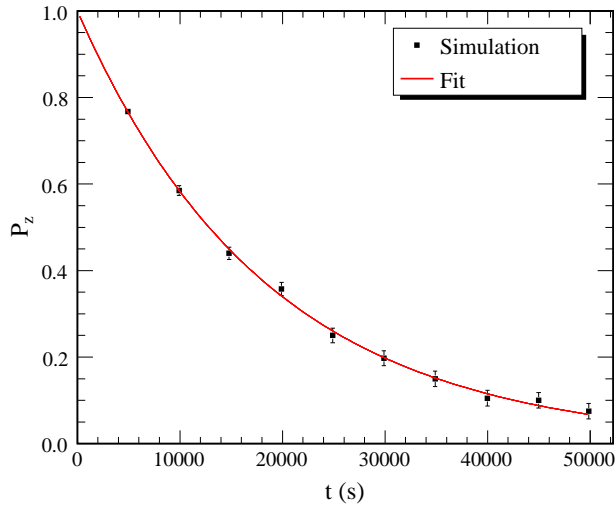}

\caption{\label{fig:expDecay}(Color online) Simulation results of the longitudinal
polarization $P_{z}$ for $^{3}$He sample in a thermal bath at $T=100\,\mathrm{mK}$
vs. time. An exponential fit to the simulation data is also shown.
The fit yields the relaxation time $T_{1}$.}

\end{figure}
We have simulated the spin relaxation of UCN and $^{3}$He atoms due
to the $v\times E$ effect for uniform and static magnetic and electric
fields. Our studies are based on the precession of classical spins
in a magnetic field. Neutrons (and $^{3}\mathrm{He}$ atoms) are confined
inside the rectangular \emph{nEDM} measurement cell of dimensions
$\left(\Delta x=10.2\,\mathrm{cm},\,\Delta y=50\,\mathrm{cm},\,\Delta z=7.6\,\mathrm{cm}\right)$,
where both magnetic and electric fields are oriented in the $\hat{\boldsymbol{z}}$
direction. The mean distance between wall collisions can be approximated
by $\lambda_{w}\simeq4V/A$ for a rectangular measurement cell, where
$V$ and $A$ are the volume and the surface area of the cell respectively.
For the \emph{nEDM} cell, $\lambda_{w}\simeq8\,\mathrm{cm}$, which
is comparable to the mean distance between wall collisions of $\lambda_{sim}=7.5\,\mathrm{cm}$
obtained from the simulations. As discussed in Sec. \ref{sec:He3 atoms},
$^{3}$He atoms can interact with excitations in the superfluid, so
that the mean time between collisions $\tau_{c}$ is determined both
by diffusion and wall collisions.

As we neglect the effect of gravity, changes in velocity are only
due to collisions, hence, between collisions, the total magnetic field
experienced in the rest frame of a particle stays constant. We have
used an analytic solution to the equation of motion of the spin in
a constant magnetic field\begin{equation}
\dot{\boldsymbol{\sigma}}=\gamma\boldsymbol{\sigma}\times\boldsymbol{B}\label{eq:equations-of-motion}\end{equation}
to calculate the spin precession between collisions. The uniform magnetic
field solution to Eq. \ref{eq:equations-of-motion} can be found to
be:\begin{equation}
\begin{array}{rcl}
\boldsymbol{\sigma}(t) & = & \left(\boldsymbol{\sigma_{0}}-\frac{(\boldsymbol{\sigma_{0}}\cdot\boldsymbol{B})\boldsymbol{B}}{B^{2}}\right)\cos(\omega t)\,+\,\frac{\boldsymbol{\sigma_{0}}\times\boldsymbol{B}}{B}\sin(\omega t)\,+\\
 &  & +\,\frac{(\boldsymbol{\sigma_{0}}\cdot\boldsymbol{B})\boldsymbol{B}}{B^{2}}\end{array}.\end{equation}

We have studied the case of totally diffuse wall collisions for both
UCN and $^{3}$He atoms. The diffuse wall collisions are in part responsible
for the spin relaxation. Our simulations have shown that, for the
$v\times E$ studies described in this paper, purely specular wall
collisions, in the absence of particle diffusion, do not result in
significant longitudinal relaxation of the spin polarization. In this
case, the velocity of a particle after a wall collision is highly
correlated to its initial velocity and the times between wall collisions
are repetitive, so that the motional field $B_{v}$ changes in a repeatable
way and does not, generally, allow the spins to precess far from their
initial states. We were able to determine from the simulation data
that only a small fraction $\left(<10^{-3}\right)$ of particle spins
were precessing away from their initial states for the case of specular
wall collisions. We believe that this effect occurs because these
particles are traveling in \emph{resonant} paths, where the correlation
of velocities and precession times between wall collisions is such
as to maximally rotate the spins out of alignment. Furthermore, the
walls of the measurement cells in the \emph{nEDM} experiment are not
likely to be purely specular.

In the simulations we track the spin state of each particle and calculate
the longitudinal, $T_{1}$, and transverse, $T_{2}$, relaxation times
due to the $v\times E$ effect for single speed UCN and for $^{3}$He
atoms at temperatures below $600\,\mathrm{mK}$.

In the longitudinal spin relaxation simulations, particles start with
their spins aligned with the holding field $B_{0}$. Due to the $v\times E$
effect, the ensemble of spins relaxes exponentially with a time constant
$T_{1}$, as shown in Fig. \ref{fig:expDecay}. In contrast, in the
transverse spin relaxation simulations, particles start with their
spins perpendicular to the holding field $B_{0}$ and the spins relax
with an exponential time constant $T_{2}$.

\subsection{\label{sec:He3 atoms}$^{3}$He atoms}

We have simulated the spin relaxation due to the $v\times E$ effect
in a magnetic holding field $B_{0}=10\,\mathrm{mGauss}$ and an electric
field $E_{0}=50\,\mathrm{kV/cm}$, for a sample of $^{3}$He atoms
at temperatures below $600\,\mathrm{mK}$. Unlike UCN (see Sec. \ref{sec:Ultra-Cold-Neutrons}),
whose speeds were constant in our simulations, the $^{3}$He sample
obeys a Maxwell-Boltzmann velocity distribution of the form\begin{equation}
f(v)=4\pi\left(\frac{M_{\star}}{2\pi kT}\right)^{3/2}v^{2}e^{-M_{\star}\, v^{2}/2kT}\,\mathrm{d}v,\label{eq:Maxwell-Boltzmann}\end{equation}
where $T$ is the temperature of the sample, $k$ is the Boltzmann
constant, and $M_{\star}\simeq2.2\, M_{3}$ is the effective mass
of $^{3}$He atoms in superfluid $^{4}$He, with $M_{3}$ as the $^{3}$He
atomic mass\citep{Krotscheck3HeMass}. The mean speed of $^{3}$He
atoms is $\bar{v}=\left\langle v\right\rangle =\sqrt{8\, kT/\pi\, M_{\star}}$,
while the root mean square speed is $v_{rms}\equiv\sqrt{\left\langle v^{2}\right\rangle }=\sqrt{3\, kT/M_{\star}}$. 

The collisions with the walls and interactions with the superfluid
were implemented to be totally diffuse. In addition, we included thermalization
of $^{3}$He atoms, so that the magnitude of the velocity is randomly
sampled from the original velocity distribution (Eq. \ref{eq:Maxwell-Boltzmann})
after each collision with either phonons or the walls of the measurement
cell.

\begin{figure}
\includegraphics{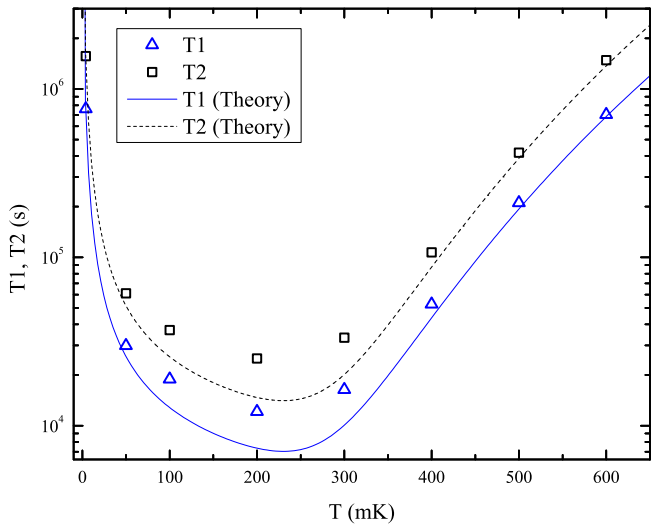}

\caption{\label{fig:He3 T1 and T2}(Color online) The spin relaxation times
$T_{1}$ and $T_{2}$ for $^{3}\mathrm{He}$ atoms as a function of
temperature, plotted together with the theoretical formulae derived
from Eq. \ref{eq:McGregorT1}.}

\end{figure}

\begin{table}
\begin{centering}
\begin{tabular}{|c|c|c|c|c|}
\hline 
T (mK) & $\bar{v}\,\mathrm{(m/s)}$ & $\lambda_{p}\,\mathrm{(cm)}$ & $\omega_{0}\tau_{D}$ & $\omega_{0}\tau_{c}$\tabularnewline
\hline
\hline 
4 & 3.573 & $\gg10^{3}$ & 4.45 & 4.45\tabularnewline
\hline 
50 & 12.63 & $\gg10^{3}$ & 1.26 & 1.26\tabularnewline
\hline 
100 & 17.86 & $\gg10^{3}$ & .890 & .890\tabularnewline
\hline 
200 & 25.26 & 126 & .668 & .593\tabularnewline
\hline 
300 & 30.94 & 6.02 & 1.18 & .224\tabularnewline
\hline 
400 & 35.73 & .696 & 5.43 & $3.65\times10^{-2}$\tabularnewline
\hline 
500 & 39.94 & .131 & 24.2 & $6.55\times10^{-3}$\tabularnewline
\hline 
600 & 43.76 & $3.33\times10^{-2}$ & 85.5 & $1.54\times10^{-3}$\tabularnewline
\hline
\end{tabular}
\par\end{centering}

\caption{\label{tab:temperature}Temperature dependence of velocity and mean
free path for a sample of $^{3}$He atoms at selected temperatures.
The $4\,\mathrm{mK}$ point represents $^{3}$He velocities comparable
to UCN.}

\end{table}
The $^{3}$He atoms can interact with excitations in the superfluid
$^{4}$He in the \emph{nEDM} cell. Due to the low fractional density
of $^{3}$He atoms, typically on the order of one $^{3}$He atom per
$10^{10}$ atoms of $^{4}$He, collisions with other $^{3}$He atoms
are negligible. The mass diffusion coefficient for $^{3}$He atoms
in superfluid He depends sensitively on the sample temperature and
was found experimentally\citep{0295-5075-58-5-718} to be \begin{equation}
D=\frac{\left(1.6\pm0.2\right)\,\mathrm{cm^{2}\cdot s^{-1}}}{(T\left[\mathrm{K}\right])^{7}}.\end{equation}

Since the phonon velocity in superfluid $^{4}$He is much higher than
the velocity of individual atoms\citep{geometricPhase}, the mean
time between phonon interactions with $^{3}$He atoms in the measurement
cell is constant for a given temperature and independent of the instantaneous
$^{3}$He velocity and can be expressed as\citep{geometricPhase}
\begin{equation}
\tau_{p}=\frac{3\, D}{v_{rms}^{2}}.\end{equation}
The effective mean time between collisions $\tau_{c}$, which takes
into account both superfluid excitations and wall collisions, is \begin{equation}
\frac{1}{\tau_{c}}=\frac{\bar{v}}{\lambda_{w}}+\frac{1}{\tau_{p}}.\label{eq:tau_c}\end{equation}

The mean time $\tau_{D}$ that it takes for $^{3}$He atoms to diffuse
to the walls of the measurement cell is useful for defining the regime
in which the formalism described by Eq. \ref{eq:v_autocorrelation}
is valid, and can be expressed as \begin{equation}
\tau_{D}=\frac{R^{2}}{\tau_{c}\, v_{rms}^{2}},\label{eq:tau_D}\end{equation}
where $R\simeq8\,\mathrm{cm}$ is the characteristic size of the measurement
cell, or the radius for a spherical cell. 

Table \ref{tab:temperature} shows the mean distance between interactions
with excitations in the superfluid $\lambda_{p}=\tau_{p}\,\bar{v}$,
for selected simulation temperatures, along with other relevant quantities.
In the low temperature regime $(T<200\,\mathrm{mK})$ the mean distance
between collisions with excitations $\lambda_{p}$ is longer than
the dimensions of the \emph{nEDM} measurement cell, so wall collisions
dominate and the mean free path is the mean distance between wall
collisions. In the high temperature regime $(T>300\,\mathrm{mK})$
$\lambda_{p}\lesssim5\,\mathrm{cm}$, so collisions with excitations
dominate. When the collision frequency is high relative to the precession
frequency, the spin relaxation times due to the $v\times E$ effect
increase. In the mid-temperature regime $\left(T\sim100-300\,\mathrm{mK}\right)$,
where the quantity $\omega_{0}\tau_{D}$ is of order unity, Eq. \ref{eq:spectrum_derived}
is not a good approximation for the frequency spectrum, and this is
where we see the largest deviation between the simulations and the
predicted relaxation times. For $\omega_{0}\tau_{D}\ll1$, a full
diffusion theory calculation is necessary to determine the frequency
spectrum. In any case, the \emph{nEDM} geometry and fields configuration
is such that the quantity $\omega_{0}\tau_{D}\gtrsim1$ for all ranges
of temperatures in our study, keeping the deviation small.

We analyzed the data from our simulations to calculate the spin relaxation
times $T_{1}$ and $T_{2}$ for $^{3}$He atoms for temperatures below
$600\,\mathrm{mK}$. The results are shown in Fig. \ref{fig:He3 T1 and T2},
where relaxation times are plotted as a function of temperature. The
simulation data agree with the theoretical diffusion calculations,
especially for temperatures higher than $400\,\mathrm{mK}$, in which
the collision frequency $1/\tau_{c}$ is large and the diffusion times
$\tau_{D}$ are long. The simulation data also agree with the relation
$T_{2}=2T_{1}$.

Our calculations and simulations have shown that the longitudinal
spin relaxation times $T_{1}$ for $^{3}$He atoms at temperatures
in the range $100-400\,\mathrm{mK}$ can be as short as $10^{4}\,\mathrm{s}$.
These relaxation times are comparable to the measurement time of \emph{nEDM}
and may be an important component of the total relaxation rate. In
this regard, the temperature is a parameter of the measurement that
can be chosen to optimize spin relaxation times.

\subsection{\label{sec:Ultra-Cold-Neutrons}Ultra Cold Neutrons}

\begin{figure}
\includegraphics{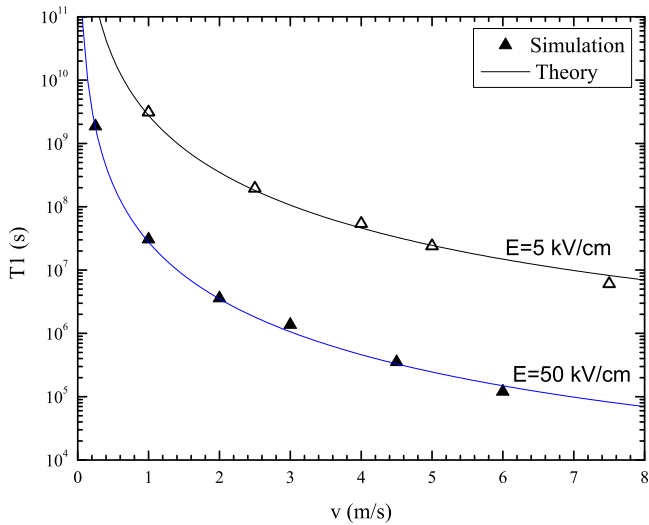}

\caption{\label{fig:UCN-agree}(Color online) Simulation results for longitudinal
spin relaxation times $T_{1}$ for UCN as a function of velocity and
with different electric field strengths. The simulation data agree
with the theoretical formalism described by Eq. \ref{eq:Theory}.}

\end{figure}

We have simulated the spin relaxation due to the $v\times E$ effect
for UCN for the conditions in the \emph{nEDM} experiment. The measurement
cell, as designed for the nEDM experiment, has a UCN Fermi potential
of $134\,\mathrm{neV}$, which is equivalent to a maximum UCN velocity
of $3.6\,\mathrm{m/s}$ and constitutes an upper limit for the velocity
spectrum of UCN. 

The relaxation times are found to be shorter for higher UCN speeds,
as discussed in Sec. \ref{sub:Theory-UCN}. We have simulated UCN
at different velocities (see Fig. \ref{fig:UCN-agree}), experiencing
diffuse collisions with the walls of the cell. The UCN do not interact
with other particles in the cell so their speed stays constant in
our simulations. The \emph{nEDM} magnetic field configuration was
used, with $B_{0}=10\,\mathrm{mGauss}$.

The results from the UCN simulations agree with the theoretical results
(Eq. \ref{eq:Theory}) in the regime in which collision frequencies
are low with respect to the Larmor frequency. The expected $1/v^{3}$
dependence of $T_{1}$ for UCN can be seen in Fig. \ref{fig:UCN-agree},
for two values of $E=5\,\mathrm{kV\cdot cm^{-1}}$ and $50\,\mathrm{kV\cdot cm^{-1}}$.
The relevant relaxation times for UCN with speeds less than $4\,\mathrm{m/s}$
are on the order of $10^{5}\,\mathrm{s}$, for the configuration of
the \emph{nEDM} measurement. These relaxation times are much longer
than the neutron lifetime and should not constitute a limiting factor
for the experiment.

\section{Conclusion}

We have studied the longitudinal and transverse spin-relaxation times
$T_{1}$ and $T_{2}$ from motional $v\times E$ fields. The results
from Monte Carlo simulations and theoretical calculations of the relaxation
times are shown to be in agreement. We investigated this phenomenon
within the context of the operating parameters for an upcoming search
for the neutron electric dipole moment utilizing Ultra Cold Neutrons
and a polarized $^{3}$He co-magnetometer. For UCN, the relaxation
times were found to be on the order of $10^{6}\,\mathrm{s}$, much
longer than the neutron lifetime. In contrast, our calculations have
shown that the relaxation times for $^{3}$He atoms will vary strongly
with temperature, and can be as low as $10^{4}\,\mathrm{s}$. Thus,
consideration of the relaxation effects from $v\times E$ motional
magnetic fields can influence the choice of the experiment's operating
temperature.

\section{Acknowledgments}

We would like to thank M. Hayden for helpful discussions. The work
in this paper was supported in part by the U.S. National Science Foundation,
grant number PHY-0555674.

\bibliographystyle{apsrev}
\bibliography{vxe}

\end{document}